\documentclass{elsart} \usepackage{graphicx} \usepackage{amssymb}
\usepackage{amsmath}
\begin{document}
\newcommand{\affinfnuniA}[3]{Dipartimento di Fisica dell'Universit\`a #1 \\ and INFN Sezione di #2, #3, Italy}
\newcommand{\affinfnuniB}[3]{Dipartimento di Scienze Fisiche dell'Universit\`a #1 \\ and INFN Sezione di #2, #3, Italy}
\newcommand{\affinfnuniC}[3]{Dipartimento di Fisica dell'Universit\`a #1 \\ and INFN gruppo collegato di #2, #3, Italy}
\newcommand{\affuni}[2]{Dipartimento di Fisica dell'Universit\`a #1, #2, Italy.}
\newcommand{\affinfn}[2]{INFN Sezione di #1, #2, Italy.}
\newcommand{\dafne}	{DA$\Phi$NE }
\newcommand{\phippp}	{\phi \rightarrow \pi^+ \pi^- \pi^0}
\newcommand{\phietag}	{\phi \rightarrow \eta \gamma}

\newcommand{\pp}	{\pi^+ \pi^- } \newcommand{\ee}	{e^+ e^- }
\newcommand{\etap}{\eta^{\prime}} \newcommand{\etapppg}{\eta^{\prime}
  \rightarrow \pi^+ \pi^- \gamma} \newcommand{\etappg}{\eta
  \rightarrow \pi^+ \pi^- \gamma} \newcommand{\etappp}	{\eta
  \rightarrow \pi^+ \pi^- \pi^0}

\newcommand{\bharad}	{e^+ e^- \to e^+ e^- (\gamma)}
\newcommand{\eephietag}	{e^+ e^- \to \phi \to \eta \gamma}

\begin{frontmatter}

\title{Measurement of \mathversion{bold}$\Gamma(\eta \to
  \pi^+\pi^-\gamma)/\Gamma(\eta \to
  \pi^+\pi^-\pi^0)$\mathversion{normal} with the KLOE Detector}
  
\collab{The KLOE / KLOE-2 Collaboration} 
\author[Frascati]{D.~Babusci},
\author[Roma2,INFNRoma2]{D.~Badoni},
\author[Cracow]{I.~Balwierz-Pytko},
\author[Frascati]{G.~Bencivenni},
\author[Roma1,INFNRoma1]{C.~Bini},
\author[Frascati]{C.~Bloise},
\author[INFNRoma1]{V.~Bocci},
\author[Frascati]{F.~Bossi},
\author[INFNRoma3]{P.~Branchini},
\author[Roma3,INFNRoma3]{A.~Budano},
\author[Uppsala]{L.~Caldeira~Balkest\aa hl},
\author[Frascati]{G.~Capon},
\author[Roma3,INFNRoma3]{F.~Ceradini},
\author[Frascati]{P.~Ciambrone},
\author[Frascati]{E.~Czerwi\'nski},
\author[Frascati]{E.~Dan\'e},
\author[Frascati]{E.~De~Lucia},
\author[INFNBari]{G.~De~Robertis},
\author[Roma1,INFNRoma1]{A.~De~Santis},
\author[Frascati]{P.~De~Simone},
\author[Roma1,INFNRoma1]{A.~Di~Domenico},
\author[INFNNapoli]{C.~Di~Donato\corauthref{cor}},
\ead{camilla.didonato@na.infn.it}
\author[Roma3,INFNRoma3]{B.~Di~Micco},
\author[INFNRoma2]{R.~Di~Salvo},
\author[Frascati]{D.~Domenici},
\author[Bari,INFNBari]{O.~Erriquez},
\author[Bari,INFNBari]{G.~Fanizzi},
\author[Roma2,INFNRoma2]{A.~Fantini},
\author[Frascati]{G.~Felici},
\author[Roma1,INFNRoma1]{S.~Fiore},
\author[Roma1,INFNRoma1]{P.~Franzini},
\author[Roma1,INFNRoma1]{P.~Gauzzi},
\author[Messina,INFNCatania]{G.~Giardina},
\author[Frascati]{S.~Giovannella},
\author[Roma2,INFNRoma2]{F.~Gonnella},
\author[INFNRoma3]{E.~Graziani},
\author[Frascati]{F.~Happacher},
\author[Uppsala]{B.~H\"oistad},
\author[Frascati]{L.~Iafolla},
\author[Uppsala]{M.~Jacewicz\corauthref{cor}},
\ead{marek.jacewicz@physics.uu.se} \corauth[cor]{Corresponding
  author.}
\author[Uppsala]{T.~Johansson},
\author[Uppsala]{A.~Kupsc},
\author[Frascati,StonyBrook]{J.~Lee-Franzini},
\author[Frascati]{B.~Leverington},
\author[INFNBari]{F.~Loddo},
\author[Roma3,INFNRoma3]{S.~Loffredo},
\author[Messina,INFNCatania,CentroCatania]{G.~Mandaglio},
\author[Moscow]{M.~Martemianov},
\author[Frascati,Marconi]{M.~Martini},
\author[Roma2,INFNRoma2]{M.~Mascolo},
\author[Roma2,INFNRoma2]{R.~Messi},
\author[Frascati]{S.~Miscetti},
\author[Frascati]{G.~Morello},
\author[INFNRoma2]{D.~Moricciani},
\author[Cracow]{P.~Moskal},
\author[Roma3,INFNRoma3]{F.~Nguyen},
\author[INFNRoma3]{A.~Passeri},
\author[Energetica,Frascati]{V.~Patera},
\author[Roma3,INFNRoma3]{I.~Prado~Longhi},
\author[INFNBari]{A.~Ranieri},
\author[Uppsala]{C.~F.~Redmer},
\author[Frascati]{P.~Santangelo},
\author[Frascati]{I.~Sarra},
\author[Calabria,INFNCalabria]{M.~Schioppa},
\author[Frascati]{B.~Sciascia},
\author[Cracow]{M.~Silarski},
\author[Roma3,INFNRoma3]{C.~Taccini},
\author[INFNRoma3]{L.~Tortora},
\author[Frascati]{G.~Venanzoni},
\author[Frascati,CERN]{R.~Versaci},
\author[Warsaw]{W.~Wi\'slicki},
\author[Uppsala]{M.~Wolke},
\author[Frascati,Beijing]{G.~Xu}
\author[Cracow]{J.~Zdebik}
\address[Bari]{\affuni{di Bari}{Bari}}
\address[INFNBari]{\affinfn{Bari}{Bari}}
\address[Beijing]{Institute of High Energy 
Physics of Academica Sinica,  Beijing, China.}
\address[CentroCatania]{Centro Siciliano di Fisica Nucleare e Struttura della Materia, Catania, Italy.}
\address[INFNCatania]{\affinfn{Catania}{Catania}}
\address[Calabria]{\affuni{della Calabria}{Cosenza}}
\address[INFNCalabria]{INFN Gruppo collegato di Cosenza, Cosenza, Italy.}
\address[Cracow]{Institute of Physics, Jagiellonian University, Cracow, Poland.}
\address[Frascati]{Laboratori Nazionali di Frascati dell'INFN, Frascati, Italy.}
\address[Messina]{Dipartimento di Fisica e Scienze della Terra dell'Universit\`a di Messina, Messina, Italy.}
\address[Moscow]{Institute for Theoretical and Experimental Physics (ITEP), Moscow, Russia.}
\address[INFNNapoli]{\affinfn{Napoli}{Napoli}}
\address[Energetica]{Dipartimento di Scienze di Base ed Applicate per l'Ingegneria dell'Universit\`a 
``Sapienza'', Roma, Italy.}
\address[Marconi]{Dipartimento di Scienze e Tecnologie applicate, Universit\`a ``Guglielmo Marconi", Roma, Italy.}
\address[Roma1]{\affuni{``Sapienza''}{Roma}}
\address[INFNRoma1]{\affinfn{Roma}{Roma}}
\address[Roma2]{\affuni{``Tor Vergata''}{Roma}}
\address[INFNRoma2]{\affinfn{Roma Tor Vergata}{Roma}}
\address[Roma3]{\affuni{``Roma Tre''}{Roma}}
\address[INFNRoma3]{\affinfn{Roma Tre}{Roma}}
\address[StonyBrook]{Physics Department, State University of New 
York at Stony Brook, USA.}
\address[Uppsala]{Department of Physics and Astronomy, Uppsala University, Uppsala, Sweden.}
\address[Warsaw]{National Centre for Nuclear Research, Warsaw, Poland.}
\address[CERN]{Present Address: CERN, CH-1211 Geneva 23, Switzerland.}

\begin{abstract}
The ratio  $R_{\eta}=\Gamma(\eta \to \pi^+\pi^-\gamma)/\Gamma(\eta \to \pi^+\pi^-\pi^0)$ 
has been measured by analysing 22 million
$\phi \to \eta \gamma$ decays collected by the KLOE
experiment at DA$\Phi$NE, corresponding to an
integrated luminosity of 558 pb$^{-1}$. The $\eta \to
\pi^+\pi^-\gamma$ proceeds both via the $\rho$ resonant
contribution, and possibly a non-resonant direct
term, connected to the box anomaly. Our result, $R_{\eta}= 0.1856\pm
0.0005_{\text {stat}} \pm 0.0028_{\text {syst}}$, points out a sizable contribution of
the direct term to the total width.
The di-pion invariant mass for the $\eta \to \pi^+\pi^-\gamma$ decay
could be described in a model-independent approach in terms of a
single free parameter, $\alpha$. The determined value of the
parameter $\alpha$ is  
$\alpha = (1.32 \pm 0.08_{\text {stat}}$$^{+0.10} _{-0.09}$$_{\text {syst}}$$\pm 0.02_{\text {theo}})$ 
GeV$^{-2}$.

\end{abstract}
\end{frontmatter}
\section{Introduction}
\label{sec:introduction}
The Chiral Perturbation Theory (ChPT) provides an accurate description of
interactions and decays of light mesons \cite{GasserLut}. 
The Wess-Zumino-Witten (WZW)
term in the ChPT Lagrangian accounts for anomalous decays 
involving an odd number of pseudoscalar mesons. 
The triangle anomaly is responsible for the two-photon decays of the 
$\pi^0 / \eta / \etap$  mesons. Both triangle and  box anomalies should 
contribute to the $\eta^{(\prime)} \to \pi^+ \pi^- \gamma$ decays. 
Since the kinematic region of the decays is far from the chiral limit, 
the amplitude of the $\pi^+\pi^-$ final state interaction has to be  
properly included. The decays are therefore often described by a resonant 
contribution due to the $\rho$-meson exchange using the Vector Meson 
Dominance (VMD) model, and an additional contact term (CT), whose strength 
is constrained by the requirement to obtain a total contribution consistent 
with the WZW term in the chiral limit. 
In the case of $\eta \to \pi^+ \pi^- \gamma$ the resonant $\rho$ contribution 
is sub-dominant, making the partial decay width sensitive to the CT, while 
for the $\eta^{\prime} \to \pi^+ \pi^- \gamma$ decay the partial width is 
dominated by the resonance but  the direct term will influence the shape of 
the di-pion invariant mass distribution. The present world average of the 
$\eta \to \pi^+ \pi^- \gamma$ partial width, $\Gamma(\eta \to \pi^+ \pi^- \gamma)$ 
= (60 $\pm$ 4)~eV \cite{PDG12}, provides strong evidence of the CT in the box 
anomaly when compared with the values obtained with and without the direct term, 
(56.3 $\pm$ 1.7)~eV and (100.9 $\pm$ 2.8)~eV, respectively \cite{Ben2003}.

Various approaches have been used to describe the final state interaction 
in these decays: the Hidden Local Symmetry (HLS) model \cite{Ben2003}, 
the chiral unitary approach \cite{Borasoy} and the Omnes function 
encoding pion-pion interaction \cite{Venugopal}. 
A model-independent approach, based on a combination of ChPT and dispersion 
theory, has been recently proposed, where a parametrisation of the 
experimental pion vector form factor is used instead of VMD \cite{Stoll}.\\
Recently, CLEO \cite{Lopez07} has measured the ratio
$R_{\eta}=\Gamma(\eta \to \pi^+ \pi^- \gamma)/\Gamma(\eta \to \pi^+
\pi^- \pi^0)= 0.175\pm 0.007_{\text {stat}} \pm 0.006_{\text {syst}}$, which differs 
by more than $3\sigma$ from the average of previous 
measurements \cite{Gormley,Thaler}, $R_{\eta}= 0.207\pm 0.004$ \cite{PDG06}.
We present a new measurement of $R_{\eta}$ with smaller statistical and systematic 
errors, together with the fit of the $M_{\pi\pi}$ distribution according to the 
model-independent approach presented in \cite{Stoll}.
\section{The KLOE detector at DA$\Phi$NE}
\label{sec:detector}
The KLOE experiment operated at the Frascati $\phi$-factory, 
DA$\Phi$NE, an $e^+e^-$ collider running at a center-of-mass energy 
of $\sim 1020$~MeV, the mass of the $\phi$ meson.
The beams collide at a crossing angle of ($\pi$ - 0.025) rad, 
producing $\phi$ mesons with a small momentum in the horizontal 
plane, $p_{\phi} = 12.5$ MeV.
The detector consists of a large cylindrical Drift Chamber (DC),
surrounded by a lead-scintillating fiber electromagnetic calorimeter (EMC) and
a superconducting coil around the EMC providing a 0.52~T field.
The DC~\cite{DCH}, 4~m in diameter and 3.3~m long, has 12,582
all-stereo tungsten sense wires and 37,746 aluminum field wires.
The chamber shell is made of carbon fiber-epoxy composite with an
internal wall of 1.1 mm thickness, the gas used is a 90\% helium,
10\% isobutane mixture. 
The spatial resolutions are $\sigma_{xy} \sim 150\ \mu$m and 
$\sigma_z \sim$~2 mm and the momentum resolution is 
$\sigma(p_{\perp})/p_{\perp}\approx 0.4\%$.
The EMC~\cite{EMC} consists of a barrel and two endcaps, for a
total of 88 modules, and covers 98\% of the solid angle. 
The modules are read out at both ends by photomultipliers, both in
amplitude and time. 
The readout granularity is $\sim$\,(4.4 $\times$ 4.4)~cm$^2$, for a total
of 2440 cells arranged in five layers. 
The energy deposits are obtained from the signal amplitude, while particle position
along fiber direction is obtained from the arrival time difference of the signals
to the photo-multipliers at the ends of each calorimeter cell.
Signals of calorimeter cells close in time and space are grouped into clusters and
the cluster energy $E$ is the sum of the cell energies.
The cluster time $T$ and position $\vec{R}$ are energy-weighted averages. 
Energy and time resolutions are, $\sigma_E/E = 5.7\%/\sqrt{E\ {\rm(GeV)}}$, 
and $\sigma_t = 57\ {\rm ps}/\sqrt{E\ {\rm(GeV)}} \oplus100\ {\rm ps}$, 
respectively.
The trigger \cite{TRG} uses both calorimeter and chamber information.
Data are then analysed by an event classification filter (EVCL), that 
organises data in different output files, according to their particle 
content \cite{NIMOffline}.
\section{Event selection}
\label{sec:eventselection}
The analysis has been performed using 558 pb$^{-1}$,
collected at $\sqrt{s} \simeq 1020$ MeV, which correspond to
about $22 \times 10^{6}\ \eta$-mesons produced. 
KLOE Monte Carlo (MC) program \cite{NIMOffline} is used to simulate 
the final states produced in $e^+e^-$ collisions, taking into account machine 
parameters and beam-related background on run-by-run basis. 
At KLOE, the $\eta$ mesons are produced together with a monochromatic 
recoil photon, $E_{\gamma}=363$ MeV, through the radiative decay $\phietag$.
The final state under study is $\pi^+\pi^-\gamma\gamma$ with the main background 
coming from $\phi \to \pi^+\pi^-\pi^0, \pi^0 \to \gamma \gamma$.  
Another important background is $\eta$ decay $\phi \to \eta \gamma\to \pi^+\pi^-\pi^0 
\gamma \to \pi^+ \pi^- 3\gamma$ with one undetected photon.
In the MC generator the signal is simulated using a matrix element
\begin{equation*}
|M|^2\simeq k^2 \sin^2\theta \left(\frac{M_{\pi\pi}}{q}\right)
\frac{\Gamma}{(M_{\rho}^2-M_{\pi\pi}^2)^2+M_{\rho}^2\Gamma^2}
\end{equation*}
where, $k$ is the photon momentum in the $\eta$ rest frame,
$\theta$ is the angle between the $\pi^+$ and the photon in the di-pion rest frame,
$q$ is the momentum of both pions in the di-pion rest frame and
$\Gamma=$124$\cdot( q/q_0)^3$ MeV with $q_0$ being the value of $q$
at $\rho$-meson resonance \cite{Gormley}.\\
After the EVCL filter, a preselection is performed, 
requiring at least two tracks with opposite charge pointing to the interaction
point (IP) and at least two clusters in time\footnote{We require for each 
cluster $|T_{clu}-R_{clu}/c|< 5\sigma_{T_{clu}}$, where $T_{clu}$ is the arrival 
time at the EMC, $R_{clu}$ is the distance of the cluster from the beam 
interaction point, and $c$ is the speed of light.},
not associated to any track, having energy $E_{clu} \ge 10$ MeV and a polar 
angle in the range $(23^\circ -157^\circ)$. 
Tracks are sorted according to the distance of the point of closest approach 
from the IP. The first two tracks with opposite charge are selected as pion candidates.
\subsection{\bf $\eta \to \pi^+ \pi^- \gamma$ selection}
\label{signal}
\begin{figure}[htb] 
\centering
\includegraphics*[width=0.65\textwidth]{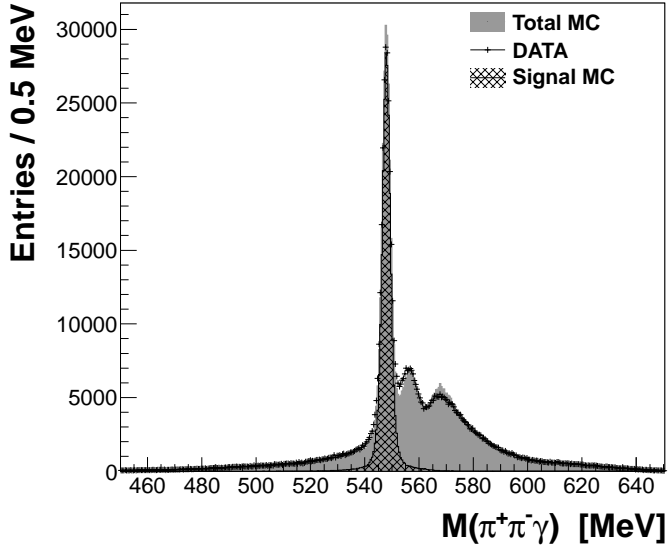}
\caption{The $\pi^+ \pi^- \gamma_{\eta}$ invariant mass distribution. 
Crosses are experimental points, the hashed area is the MC signal $\etappg$, 
the filled region represents the total MC.
 Relevant background is due to $\phi\to\pi^+\pi^-\pi^0$ events and much smaller contribution from $\phi$-meson decay into kaons for higher masses as well as $\phi\to\eta\gamma$ events for the masses below the signal peak.}
\label{pppresults1}
\end{figure}

We require that the most energetic cluster has an energy
$E_{clu}>250$ MeV and we identify it as the photon ($\gamma_{\phi}$) recoiling against
the $\eta$ in the $\phi \to \eta \gamma$ decay. Moreover, we ask that the
$\gamma_{\phi}$ is inside the calorimeter barrel (with a polar angle in the 
range $55^{\circ}-125^{\circ}$), 
to reject events with cluster split between barrel and endcap.
To reject electrons, cuts on cluster-track association and identification 
by time of flight (TOF) are used. 
These cuts reject Bhabha scattering background and other processes with
electrons in the final state.
We exploit the $\phi \to \eta \gamma$ decay kinematics, to evaluate the
$\gamma_{\phi}$ energy:
\[
\vec{p}_{\phi}=\vec{p}_{\eta}+\vec{p}_{\gamma_{\phi}} \qquad
E_{\gamma_{\phi}}=
\frac{M_{\phi}^2-M_{\eta}^2}{2(E_{\phi}-|\vec{p}_{\phi}|\cos\vartheta)}
\] 
where, $\vartheta$ is the angle between $\gamma_{\phi}$ and the 
$\phi$ meson momentum, $\vec{p}_{\phi}$, measured run by run with high 
accuracy using Bhabha scattering events. 
This allows us to improve the energy measurement accuracy of
the recoil photon to $0.1\%$. 
Using $\phi$ and $\pi$-mesons momenta, we determine the direction 
of the photon ($\gamma_{\eta}$) from $\eta$ decay: 
\[
\vec{p}_{\gamma_{\eta}} = \vec{p}_{\phi}- \vec{p}_{\pi^+} -
\vec{p}_{\pi^-}-\vec{p}_{\gamma_{\phi}}
\]
the $\gamma_{\eta}$ photon direction is then compared with the direction of 
each neutral cluster: $\Delta \varphi = \varphi_{\mathrm{clu}} - \varphi_{\gamma_{\eta}}$
(here, and in the following, the angles are evaluated 
using variables in the transverse plane\footnote{The azimuthal angle of the 
cluster is measured with an angular resolution of 6 mrad using the position 
of the calorimeter cell. The polar angle is instead determined by the time 
difference of the signals at each side of the barrel and is affected by 
larger uncertainty. The use of azimuthal angle reduces the systematics.}). 
If no clusters with $\Delta \varphi < 8.5^{\circ}$ are found, the event 
is rejected. The cluster with the minimum value of $\Delta\varphi$ 
is identified with $\gamma_{\eta}$. 
In order to reject the $\phi \to \pi^+ \pi^- \pi^0$ background, the angle 
between the two photons in the $\pi^0$ reference frame, 
evaluated using the $\phi$ and the $\pi$-meson momenta, 
is calculated and it is required to be smaller than $165^{\circ}$. 
The $\pi^+ \pi^- \gamma$ mass spectrum is shown in Fig.~\ref{pppresults1}. 
The candidate events are selected requiring $539.5$ MeV 
$< M_{\pi^+ \pi^- \gamma} < 554.5$ MeV.
\subsection{\bf $\eta \to \pi^+ \pi^- \pi^0$ selection}
\label{sec:ppp}
\begin{figure}[htb]
\centering
\includegraphics[width=0.52\textwidth]{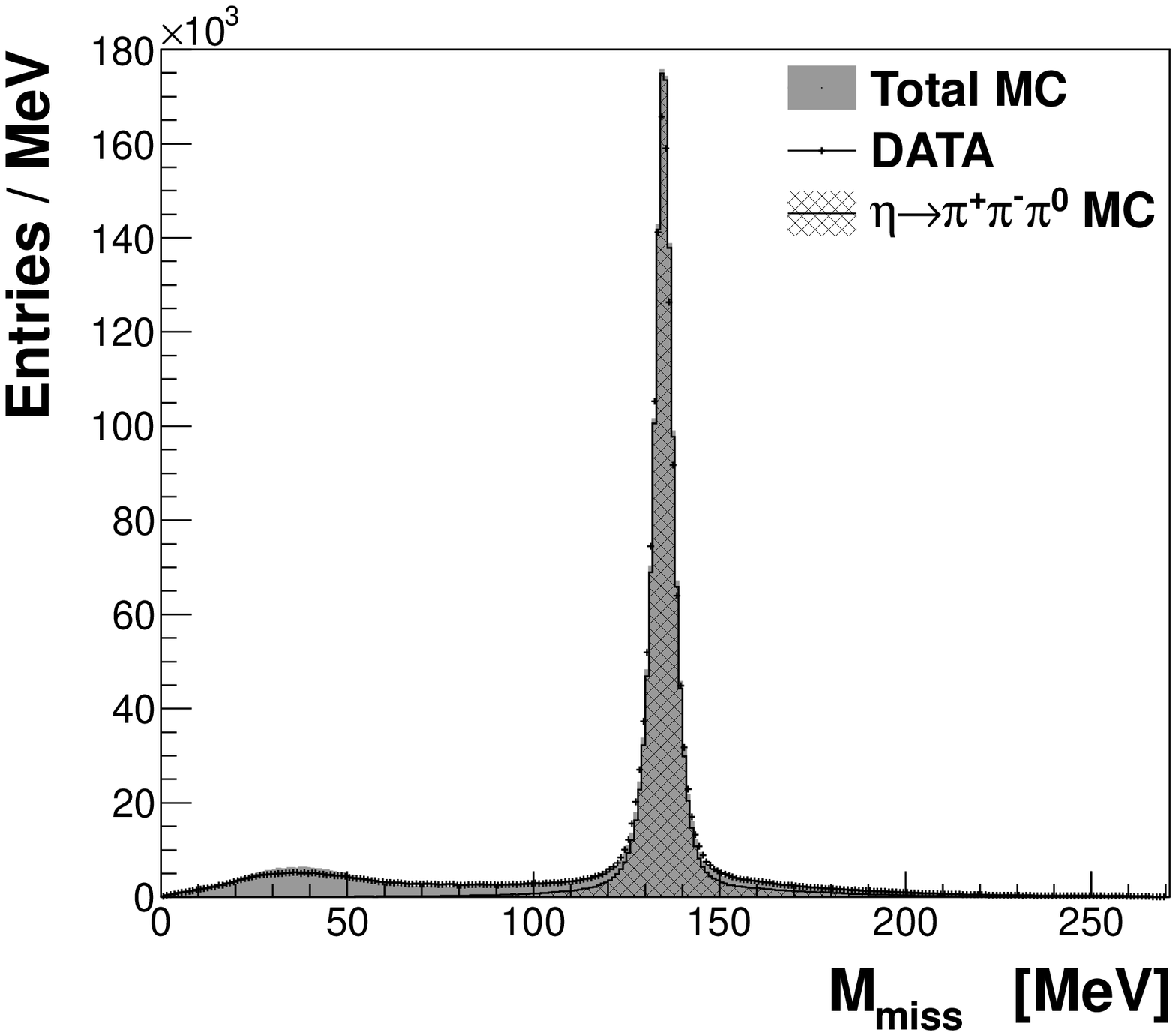}%
\includegraphics[width=0.52\textwidth]{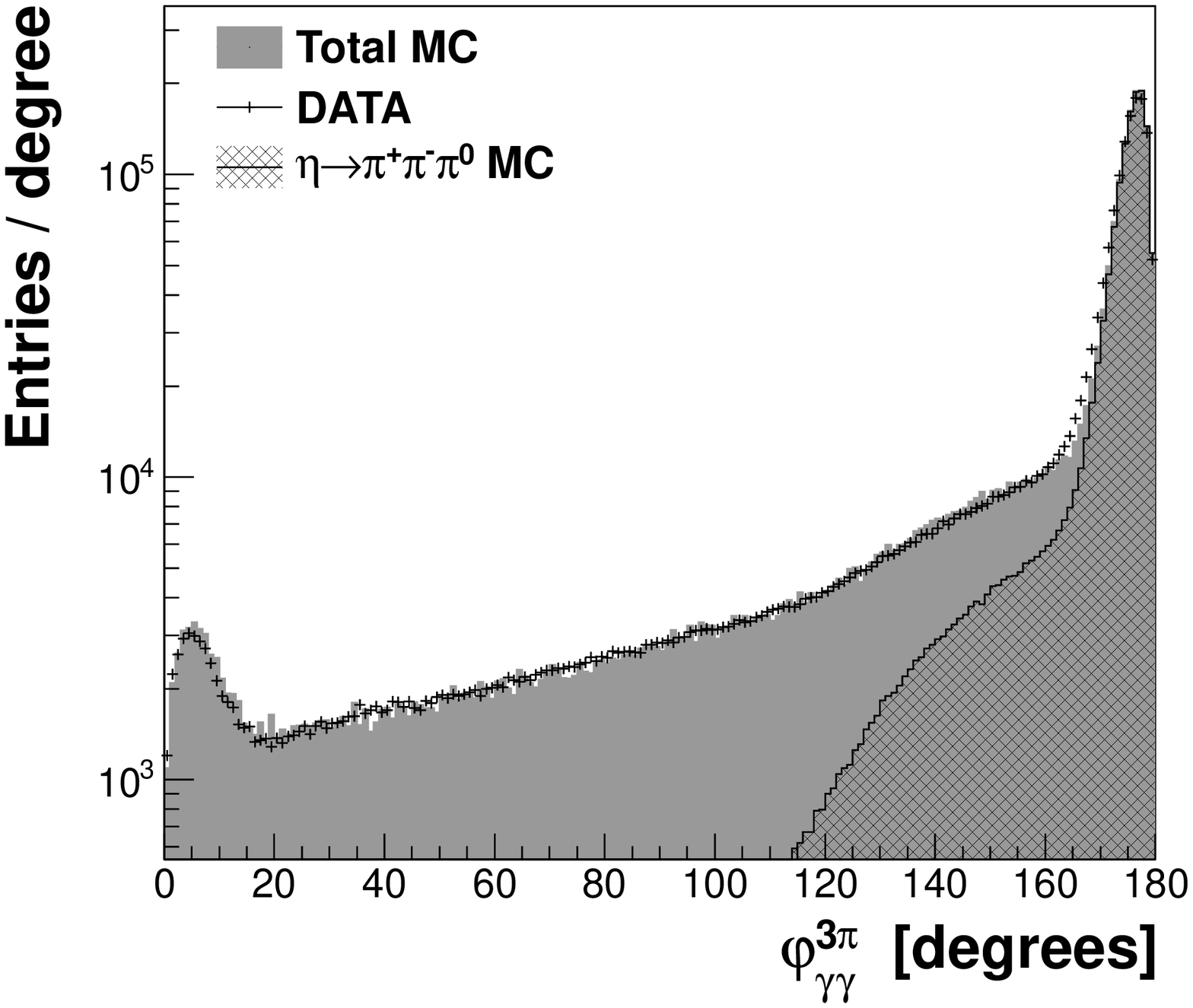}
\caption{Normalisation sample $\eta \to \pi^+ \pi^- \pi^0$.
Left - $\pi^+ \pi^- \gamma_{\phi}$ missing mass distribution. Right -
 event distribution for the angle between prompt neutral clusters in the $\pi^0$ rest frame
evaluated in the transverse plane, $\varphi_{\gamma\gamma}^{3\pi}$. 
Crosses are experimental points, the hashed area is the MC $\eta \to \pi^+ \pi^- \pi^0$, 
the filled region represents the total MC, where the only relevant background contribution is due to $\phi\to\pi^+\pi^-\pi^0$ events.}
\label{figPPP}
\end{figure}

The process $\phi \to \eta \gamma$ with $\eta \to \pi^+ \pi^- \pi^0$
represents a good control sample, having a topology similar to the signal. 
Moreover, in the ratio $\Gamma(\eta \to \pi^+ \pi^-
\gamma)/\Gamma(\eta \to \pi^+ \pi^-\pi^0)$ the luminosity, 
the $\phi$ production cross section and the $BR(\phi \to \eta \gamma)$ 
 cancel out. 
We use the same preselection as for the $\eta \to \pi^+ \pi^-
\gamma$ signal and calculate the missing four-momentum:
\[
\bf{P}_{\rm miss}=\bf{P}_{\phi}-\bf{P}_{\pi^+}-\bf{P}_{\pi^-}-\bf{P}_{\gamma_{\phi}}
\]
where the variables in the formula represent the four-momenta of the
$\phi$ meson and of the decay products. For the  $\eta \to \pi^+
\pi^- \pi^0$ sample, the missing mass peaks at
the $\pi^0$ mass value and we select events with
$|M_{\rm miss}-M_{\pi^0}|<15$ MeV. The remaining background is rejected
by an angular cut applied to the two photons
in the $\pi^0$ rest frame, $\varphi_{\gamma\gamma}^{3\pi}>165^{\circ}$. 
Figure~\ref{figPPP} shows the distribution of the missing mass 
and $\varphi_{\gamma\gamma}^{3\pi}$.
The two cuts select $N(\eta \to \pi^+\pi^-\pi^0)= 1.116 \cdot 10^3$ events.
The global selection efficiency is $\varepsilon = 0.2276\pm0.0002$ with 
residual background contamination of $0.65\%$. 

\section{Results}

\subsection{The ratio $\Gamma(\eta \to \pi^+\pi^-\gamma)/\Gamma(\eta \to \pi^+\pi^-\pi^0)$ }
\label{sec:fit-invmass}
\begin{figure}[htb]
\centering  \includegraphics*[width=0.75\textwidth]{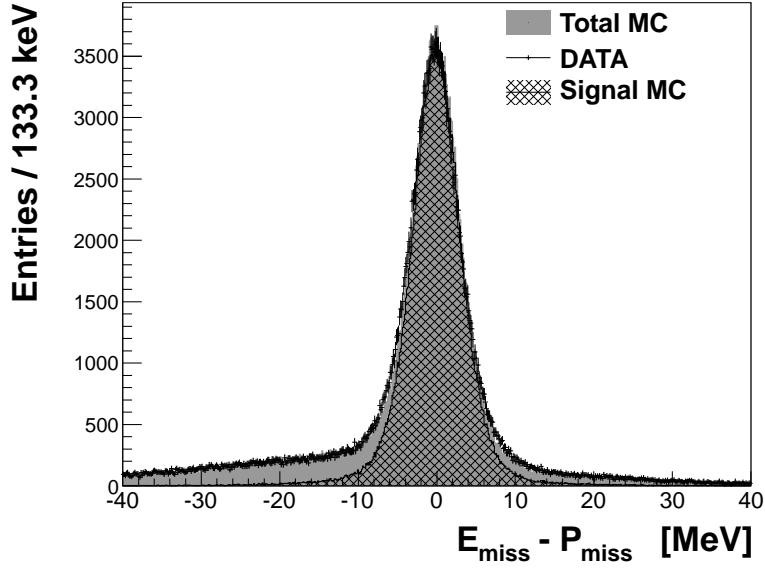}
\caption{$E_{\rm miss}-P_{\rm miss}$ distribution for the $\pi^+ \pi^-
  \gamma_{\phi}$. The sample has been selected applying all the cuts described in Sec.\ref{signal}. 
The event counting is performed in the region 
$|E_{\rm miss}-P_{\rm miss}|< 10$ MeV.}
\label{EPfit}
\end{figure}

The total selection efficiency for the signal $\eta \to \pi^+\pi^-\gamma$ 
is $\varepsilon = 0.2131 \pm 0.0004$. 
In the final sample, the relative weights of signal and background 
components are evaluated with a fit to the $E_{\rm miss}-P_{\rm miss}$ 
distribution of the $\pi^{+} \pi^{-} \gamma_{\phi}$ system, with the MC 
shapes of the remaining background and signal MC, Fig.~\ref{EPfit}. 
Signal events are counted in the range $|E_{\rm miss}-P_{\rm miss}|<10$ MeV. 
We find  $N(\eta \to \pi^+\pi^-\gamma)= 204950 \pm 497$ events, with a 
background contamination at level of $10\%$.
The analysis has been repeated on an independent 
sample selected without EVCL filter to evaluate any 
bias due to the event classification.
An overall correction factor is used to account for data/MC difference 
related to event classification: 
$K_{EVCL}= \frac{\varepsilon^{MC}_{\pi^+\pi^-\gamma} \cdot \varepsilon^{data}_{\pi^+\pi^-\pi^0} }
{\varepsilon^{data}_{\pi^+\pi^-\gamma} \cdot \varepsilon^{MC}_{\pi^+\pi^-\pi^0} }
= 1.010\pm0.009$.

Combining the results we obtain the ratio:
\begin{equation*}
R_{\eta}=\frac{\Gamma(\eta \to \pi^+ \pi^- \gamma)}{\Gamma(\eta \to
  \pi^+ \pi^- \pi^0)}= 0.1856\pm0.0005_{\text {stat}} \pm 0.0028_{\text {syst}}
\end{equation*}
This result is in agreement with the recent CLEO measurement \cite{Lopez07}, 
while improving the accuracy of a factor better than three, thus
confirming a smaller value for $R_{\eta}$ with respect to previous 
evaluations \cite{Gormley, Thaler}.

The systematic uncertainties due to analysis cuts have been evaluated
by varying the cuts and re-evaluating the value of $R_{\eta}$.
Each cut is moved $\pm 2 \sigma$ with respect to the value used in the analysis, 
where $\sigma$ is the resolution on the reconstructed variable. 
The corresponding error for each source of
systematics is reported in Table \ref{tab:systematics}. The total error is taken
as the quadratic sum of all of the contributions.
\begin{table}[!t]
 \begin{center}
  {\scriptsize
    \begin{tabular}{lllc}
    \hline Source of Uncertainty& Cut Value& Window cut  & Fractional Error on $R_{\eta}$ \\ \hline 
    $\varphi_{\gamma\gamma}^{\pi^+\pi^-\gamma}$ & $ < 165^{\circ}$ & $\pm 2^{\circ}$ & $\pm0.6\%$ \\       
    $\Delta \varphi$ & $ < 8.5^{\circ} $& $  \pm 2^{\circ}$ & $\pm 0.4\%$\\ 
    $|M_{\pi^+ \pi^-\gamma}-M_{\eta}| $&$< 7.5$ MeV &$\pm 2$ MeV & $\pm 0.6\%$\\ 
    $E_{min}^{\gamma}$&$ > 10$ MeV & $ \pm 2$ MeV  & $\pm 0.1\%$\\ 
    $E_{clu}^{\gamma_{\phi}}$ & $ > 250 $ MeV & $ \pm 4$ MeV  & $\pm 0.1\%$\\
    $|M_{\rm miss} - M_{\pi^0}|$&$ < 15 $ MeV &$ \pm 4$ MeV & $\pm 0.4\%$ \\ 
    $\varphi_{\gamma\gamma}^{3\pi}$&$ >  165^{\circ} $&$ \pm 2^{\circ}$ & $\pm 0.1\%$\\ \hline
 EVCL && & $\pm0.9\%$ \\
  Fit $E_{\rm miss}-P_{\rm miss}$ & && $\pm0.6\%$ \\ \hline
Total                    &&& $1.5\%$ \\ \hline    
    \end{tabular}}
   \caption{Summary table of systematic uncertainties on $R_{\eta}$.}
  \label{tab:systematics}
  \end{center}
\end{table}
%
\subsection{Fit to di-pion invariant mass}
\label{sec:fit-invmass}

\begin{figure}[htb]
\centering
\includegraphics[width=0.65\textwidth]{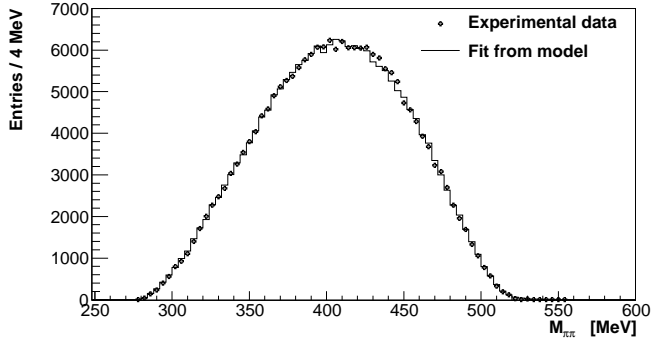}
\caption{Distribution of $M_{\pi\pi}$ after background subtraction
  (black markers). Histogram is the fit of Eq.~(\ref{eqStoll}), corrected for acceptance and 
experimental resolution.}
\label{fig:Mpp}
\end{figure}

The $M_{\pi\pi}$ dependence of the decay amplitude has been studied in several 
frameworks. The HLS model, in particular, has been investigated in \cite{Picciotto92} and
more recently in \cite{Ben2003}. 
In this approach, the relative strength of the CT 
and the resonance contribution from VMD are fixed.
The model-independent approach in \cite{Stoll}, based on ChPT and dispersive analysis,
does not fix this relative strength and parametrises the CT via a process-specific term.
We use the last method to fit the di-pion invariant mass distribution.
The function describing the partial width as a function of 
$s_{\pi\pi} = M_{\pi\pi}^2$ is the following:
\begin{equation}
 \frac{d\Gamma(\eta \to\pi^+\pi^-\gamma)}{d s_{\pi\pi}} =
 \left|AP(s_{\pi\pi})F_V(s_{\pi\pi})\right |^2
\Gamma_0(s_{\pi\pi})
\label{eqStoll}
\end{equation}
where, $A$ is a normalisation factor and
\begin{equation*}
\Gamma_0(s_{\pi\pi})=\frac{1}{3\cdot2^{11}\cdot \pi^3 M_{\eta}^3}\left (M_{\eta}^2-s_{\pi\pi}\right )^3s_{\pi\pi} \cdot \beta_{\pi}^3
\end{equation*}
is the simplest gauge-invariant matrix element multiplied by the phase-space term with $\beta_{\pi}=\sqrt{1-4M_{\pi}^2/s_{\pi\pi}}$. 
$F_V(s_{\pi\pi})$ is the pion vector form factor, approximated in the energy range of interest by 
the polynomial $|F_V(s_{\pi\pi})|=1+(2.12\pm0.01)s_{\pi\pi}+
(2.13\pm0.01)s^2_{\pi\pi}+(13.80\pm0.14)s^3_{\pi\pi}$, where $s_{\pi\pi}$ is expressed in GeV$^{2}$ \cite{Stoll}.
The $P(s_{\pi\pi})$ function, a process-specific part, 
can be treated perturbatively in the frame of
ChPT, for the decay of light mesons. Taylor expansion around $s_{\pi\pi}=0$
gives $P(s_{\pi\pi})= 1+\alpha \cdot s_{\pi\pi}+\mathcal{O}(s^2_{\pi\pi})$.
We fit the $M_{\pi\pi}$ distribution by minimising the variable:
\begin{equation}
\label{chi2}
\chi^2 = \Sigma_{i}^{Nbin}\frac{(N_{i}^{data}-\Sigma_{j}^{Nbin}N_{j}^{Teo}\varepsilon_{j}S_{ij})^2}{\sigma_{i}^{2}}
\end{equation}
where, $N_{i}^{data}$ is the content of $i-$th bin 
after background subtraction, $N_{j}^{Teo}$ is the content of $j-$th bin of the 
expected $M_{\pi\pi}$ spectrum as from Eq.(\ref{eqStoll}), 
$\varepsilon_{j}$ is the efficiency, $S_{ij}$ is the smearing matrix and 
$\sigma_i^{2} =  \sigma^2_{N_i^{data}}+\sigma^2_{N^{Teo}_i}$, with 
$\sigma^2_{N^{Teo}_i}= \Sigma_{j}^{Nbin}(N_{j}^{Teo})^2(\sigma^2_{\varepsilon_{j}}S^2_{ij}+ \varepsilon^2_{j} \sigma^2_{S_{ij}})$.
Figure~\ref{fig:Mpp} shows the measured distribution compared with
results of the fit taking into account efficiency and smearing.

Minimising the function in Eq.~(\ref{chi2}) we get 
\begin{equation*}
\alpha =\left(1.32\pm0.08_{\text {stat}} \, ^{+0.10} _{-0.09} \,_{\text {syst}} \pm 0.02_{\text {theo}}\right ) {\text GeV}^{-2}
\end{equation*}
with $\chi^2 / Ndf =61/64$. The theoretical error $0.02$~GeV$^{-2}$ accounts for uncertainty
due to vector form factor parametrisation, and is determined mainly by 
the accuracy of the existing $e^+e^- \to \pi^+\pi^-$ data.
 The fit is insensitive to the addition of a quadratic term in 
$P(s_{\pi\pi})$. The contributions to systematic uncertainty on $\alpha$ are listed in Table 
\ref{tab:systematics2}. 
\begin{table}[!t]
 \begin{center}
  {\scriptsize
    \begin{tabular}{lllc}
    \hline Source of Uncertainty& Cut Value& Window cut  & $ \Delta \alpha$ (GeV$^{-2}$)\\ \hline 
    $\varphi_{\gamma\gamma}^{\pi^+\pi^-\gamma}$ & $ < 165^{\circ}$ & $\pm 2^{\circ}$ & $+0.07/-0.03$\\       
    $\Delta \varphi$ & $ < 8.5^{\circ} $& $  \pm 2^{\circ}$ & $+0.05/-0.06$ \\ 
    $|M_{\pi^+ \pi^-\gamma}-M_{\eta}| $&$< 7.5$ MeV &$\pm 2$ MeV &$+0.04/-0.04$\\ 
    $E_{min}^{\gamma}$&$ > 10$ MeV & $ \pm 2$ MeV  &$+0.01/-0.04$\\  \hline    
Total                   &&  & $+0.10/-0.9$\\ \hline    
    \end{tabular}}
   \caption{Summary table of systematic uncertainties on $\alpha$ parameter.}
  \label{tab:systematics2}
  \end{center}
\end{table}
The value of $\alpha$ is in agreement with the result of the WASA Collaboration obtained from the
fit to the $\gamma_{\eta}$ spectrum giving 
$\alpha = (1.89 \pm 0.25_{\text {stat}} \pm 0.59_{\text {syst}}\pm 0.02_{\text {theo}})$ GeV$^{-2}$ \cite{WASA}.
\section{Conclusions}
\label{sec:conclusions}
Using a data sample corresponding to an integrated luminosity of 558
pb$^{-1}$, we select about $205000$ $\eta \to \pi^+ \pi^- \gamma$ and
$1116000$ $\etappp$ events from the $\phi \to \eta \gamma$ decays. 
We obtain the ratio of the partial widths:
\begin{equation*}
\Gamma(\eta \to \pi^+\pi^-\gamma)/\Gamma(\eta \to \pi^+\pi^-\pi^0)= 0.1856\pm0.0005_{\text {stat}} \pm 0.0028_{\text {syst}}
\end{equation*}
in agreement with the most recent result from CLEO
\cite{Lopez07}.\\ 
Combining our measurement with the world average value
$\Gamma(\eta \to \pi^+\pi^-\pi^0)= (295 \pm 16)$ eV \cite{PDG12}, we find 
$\Gamma(\eta \to \pi^+\pi^-\gamma)=(54.7 \pm 3.1)$ eV,
which is in agreement with the value expected in the HLS context including 
the contact-term contribution \cite{Ben2003}.\\ 
We have measured the di-pion invariant mass distribution and performed a fit using 
the model-independent approach of Ref. \cite{Stoll}.
The fit gives $\alpha = (1.32 \pm 0.08_{\text {stat}}$$^{+0.10} _{-0.09}$$_{\text {syst}}$$\pm 0.02_{\text {theo}})$ 
GeV$^{-2}$.
\section*{Acknowledgments}
\label{acknowledgments}
 We thank the DA$\Phi$NE team for their efforts in maintaining low 
 background
 running conditions and their collaboration during all data taking. 
 We want
 to thank our technical staff: G.F. Fortugno and F. Sborzacchi for their
 dedication in ensuring efficient operation of the KLOE computing 
 facilities;
 M. Anelli for his continuous
 attention to the gas system and detector safety; A. Balla, M. Gatta, G.
 Corradi and G. Papalino for electronics maintenance; M. Santoni, G. 
 Paoluzzi
 and R. Rosellini for general detector support; C. Piscitelli for his 
 help
 during major maintenance periods. This work was supported in part by 
 the EU
 Integrated Infrastructure Initiative Hadron Physics Project under 
 contract
 number RII3-CT- 2004-506078; by the European Commission under the 7th
 Framework Programme through the ‘Research Infrastructures’ action of 
 the
 ‘Capacities’ Programme, Call: FP7-INFRASTRUCTURES-2008-1, Grant 
 Agreement
 No. 227431; by the Polish National Science Centre through the Grants No. 
 0469/B/H03/2009/37, 0309/B/H03/2011/40, DEC-2011/03/N/ST2/02641, 
2011/01/D/ST2/00748 and by the 
 Foundation for Polish Science through the MPD programme and the 
project HOMING PLUS BIS/2011-4/3

\end{document}